\def\h{{\frac{1}{2}}}

\def\tr{{\rm tr}}
\def\q{{\frac{1}{4}}}

\documentstyle[12pt]{article}
\begin{document}
\begin{flushright}
DAMTP/94-01\\
hep-th/9401049
\end{flushright}
\vskip 0.5in
{\begin{center}
\baselineskip 22pt
{\Large \bf Dynamical r-matrices and Separation of Variables: The
Generalised Calogero-Moser Model}
\vskip 0.5in
\baselineskip 13pt
Tomasz Brzezi\'nski\footnote{Benefactors' Scholar of St. John's
College, Cambridge}
 \\[.3in]
{\em  Department of Applied Mathematics \& Theoretical
Physics \\  University of Cambridge, Cambridge  CB3 9EW, U.K.}
\vspace{24pt}

January 1994
\end{center}
\vspace{20pt}
\baselineskip 14pt

\begin{quote}ABSTRACT A generalisation of the classical Calogero-Moser
model obtained by coupling it to the Gaudin model  is considered.
The recently found classical dynamical r-matrix [E. Billey, J. Avan
and O. Babelon, PAR LPTHE 93-55]  for the Euler-Calogero-Moser model
is used to separate variables for this generalised  Calogero-Moser
model in the case in which there are two Calogero-Moser particles. The
model is then canonically quantised and the same classical r-matrix is
employed to separate variables in the Schr\"odinger equations.
\end{quote}

\vspace{20pt} }
\baselineskip 20pt

\section{Introduction}
Separation of variables in the Hamilton-Jacobi equation is one of the
methods of solving completely integrable models of classical
mechanics (see e.g.
\cite{arnold1}). If the model admits a Lax pair representation and
hence is described by a classical Yang-Baxter
algebra \cite{babelon6}, the separation of variables can be
achieved by the functional
Bethe Ansatz \cite{sklyanin1}\cite{sklyanin4}. By a classical
Yang-Baxter algebra we mean an infinite-dimensional Poisson bracket
algebra given by an $N\times N$ matrix $L(\lambda)$,  of which
components  are
dynamical variables, and an $N\sp 2\times N\sp 2$ matrix
$r\sb{12}(\lambda ,\mu)$, which may depend on dynamical variables
\cite{avan3}\cite{avan5}\cite{braden1}\cite{sklyanin5}\cite{billey1}
and is such that
\begin{equation}
\{L\sb 1(\lambda), L\sb 2(\mu)\} = [r\sb{12}(\lambda,\mu), L\sb
1(\lambda)] - [r\sb{21}(\mu ,\lambda),L\sb 2(\mu)].
\label{yang}
\end{equation}
Here $\lambda ,\mu$ are free parameters, $L\sb 1(\lambda) =
L(\lambda)\otimes I$ and $L\sb 2(\mu) =
I\otimes L(\mu)$,  and $r\sb{21}(\lambda,\mu) =
Pr\sb{12}(\lambda,\mu)P$, where $P$ is an $N\sp 2\times N\sp 2$ permutation
matrix. $L(\lambda)$ is called a Lax matrix and
$r\sb{12}(\lambda, \mu)$ is called an r-matrix.
Matrix $r\sb{12}(\lambda,\mu)$ has  to satisfy a relation necessary
to ensure that the Poisson bracket (\ref{yang}) obey the Jacobi
identity. This relation can take a form of either the classical Yang-Baxter
equation (see e.g. \cite{belavin2}) or its
dynamical generalisations
\cite{babelon6}\cite{sklyanin5}\cite{eilbeck1}  if
$r\sb{12}(\lambda,\mu)$ depends  on dynamical variables. We deal with
this last case in the present paper.

Once the classical dynamical system is written in the form
(\ref{yang}), we can define  a generating function for
integrals of  motion,
\begin{equation}
t(\lambda) = \h \tr L\sp 2(\lambda).
\label{gt}
\end{equation}
{}From (\ref{yang}) it follows that $\{t(\lambda), t(\mu)\} = 0$, hence
the integrals generated by $t(\lambda)$ are in involution and the
system is integrable. The main  goal of the functional Bethe Ansatz is
to separate the variables in the Hamilton-Jacobi equations for the
integrals generated by $t(\lambda)$.

For  models with
$sl(2)$ symmetry the functional Bethe Ansatz can be
described as follows. We begin with a Lax matrix
\begin{equation}
L(\lambda) = \pmatrix{A(\lambda) & C(\lambda) \cr B(\lambda) &
-A(\lambda)}
\label{lax}
\end{equation}
and, for simplicity,  we assume the following form of a classical r-matrix
\begin{equation}
r\sb{12}(\lambda,\mu) = \pmatrix{a(\lambda , \mu)&0&0&0\cr
0&0&b(\lambda,\mu)&0\cr 0&c(\lambda , \mu)&0&0\cr 0&0&0&a(\lambda ,
\mu)},
\label{2r}
\end{equation}
for some functions $a(\lambda , \mu)$, $b(\lambda , \mu)$ and
$c(\lambda , \mu)$. Functions $a(\lambda , \mu)$, $b(\lambda , \mu)$ and
$c(\lambda , \mu)$  satisfy relations necessary to ensure that the
Poisson bracket (\ref{yang}) obey the Jacobi identity. The r-matrices
of this type  appear in several integrable models such as homogeneous
spin chains, heavy tops and
systems on Riemannian manifolds of constant curvature. For any Lax
matrix (\ref{lax}) and any r-matrix (\ref{2r}) the Poisson bracket
algebra (\ref{yang}) takes the form
\begin{eqnarray}
\{A(\lambda) , B(\mu)\} & = &b(\lambda,\mu)B(\lambda) +
a(\mu ,\lambda)B(\mu), \nonumber \\
\{A(\lambda) , C(\mu)\} & = &-c(\lambda , \mu)C(\lambda)-
a(\mu,\lambda)C(\mu), \label{gabc}\\
\{B(\lambda) , C(\mu)\} & = &2c(\lambda ,\mu)A(\lambda)
+2b(\mu,\lambda)A(\mu), \nonumber
\end{eqnarray}
and zero for the remaining brackets. The separation variables are
defined by
\begin{equation}
B(X\sb i ) = 0, \qquad P\sb i = A(X\sb i ).
\label{px}
\end{equation}
Using the algebra (\ref{gabc}) we find \cite{kuznetsov1}
$$
\{X\sb i,X\sb j\} = \{P\sb i,P\sb j\} = 0, \qquad
\{X\sb i,P\sb j\} = \delta\sb{ij} \lim\sb{\lambda\rightarrow X\sb i}
{a(\lambda ,X\sb i)B(\lambda)\over B'(X\sb i)},
$$
so that a sufficient condition for $(X\sb i, P\sb i)$ to be canonical
variables reads
$$
\lim\sb{\lambda\rightarrow X\sb i}
{a(\lambda ,X\sb i)B(\lambda)\over B'(X\sb i)} =1,
$$
or equivalently
\begin{equation}
a(X\sb i +h,X\sb i) = {1\over h} +O(h).
\label{condition}
\end{equation}
We stress that  condition (\ref{condition}) remains the same for
systems described by non-dynamical and dynamical r-matrices. This in
particular allows one to  apply the functional Bethe Ansatz procedure
to the latter, as discovered in \cite{eilbeck1}. This observation
seems to be of great importance and  interest, since there
are integrable models, such as the Calogero-Moser model, for which the
non-dynamical r-matrices do not exist. The models of
this kind have recently attracted much attention and the theory of
dynamical r-matrices requires closer investigation \cite{sklyanin5}.

Having defined the separation variables we easily find the
separated equations
\begin{equation}
P\sb i\sp 2 = t(X\sb i),
\label{pt}
\end{equation}
which follow directly from (\ref{px}).

In this paper we employ this procedure of separation of
variables to solve the generalised Calogero-Moser model. In order to
do so we use the recently introduced r-matrix \cite{billey1} which
depends on dynamical variables. In the next section we describe a
generalisation of the Calogero-Moser model which is achieved by
coupling the Gaudin model to it. Then we specialise to the case in which
there are two Calogero-Moser particles. In this case the system has
$sl(2)$ symmetry and is governed by an r-matrix of the form
(\ref{2r}) which in addition  satisfies condition (\ref{condition}),
so that the
separation of variables procedure can be applied. We do it in Section
4. In Section 5 we quantise the model and use the quantum version of
the functional Bethe Ansatz to separate variables in the Schr\"odinger
equation \cite{kuznetsov3}. The quantum functional Bethe Ansatz we use
is obtained by
the canonical quantisation of the classical procedure just described.

\section{Description of the Model}
In \cite{billey1} Billey, Avan and Babelon proposed the
parameter dependent classical dynamical  r-matrix for the
generalisation of the
Calogero-Moser model constructed by Gibbons and Hermsen
\cite{gibbons1}. The model is governed by the Hamiltonian
\begin{equation}
H = \h\sum\sb{i=1}\sp{N} p\sb i\sp 2  -\h \sum\sb{\stackrel{i,j =1}{i\neq
j}}\sp N {f\sb{ij}f\sb{ji} \over (q\sb i-q\sb j)\sp 2},
\label{ham.calo}
\end{equation}
where the dynamical variables $(q\sb i,p\sb i)\sb{i=1,\ldots , N}$,
$(f\sb{ij})\sb{i,j=1,\ldots , N}$ satisfy the Poisson bracket algebra
\begin{equation}
\{q\sb i , q\sb j\} = \{p\sb i , p\sb j\} = 0, \quad\{q\sb i , p\sb
j\} =\delta\sb {ij}, \qquad
\{f\sb{ij} , f\sb{kl}\} = \delta\sb{jk}f\sb{il} -
\delta\sb{il}f\sb{kj} \label{commute.calo}.
\end{equation}
The model is integrable, when restricted to the surfaces $(f\sb{ii} =
{\rm const})\sb{i=1,\ldots , N}$. In this section we propose a further
generalisation of (\ref{ham.calo}) which is achieved by introducing  a
non-trivial
internal structure to the dynamical variables $(f\sb{ij})\sb{i,j=1,\ldots
, N}$. We consider a system
governed by the Hamiltonian (\ref{ham.calo}) but with the
dynamical variables $(q\sb i,p\sb i)\sb{i=1,\ldots , N}$,
$(f\sp\alpha\sb{ij})\sb{\stackrel{i,j=1,\ldots , N}{\alpha = 1,\ldots ,M}}$,
which satisfy
\begin{equation}
\{f\sp\alpha\sb{ij} , f\sp\beta\sb{kl}\} =
\delta\sb{\alpha\beta}(\delta\sb{jk}f\sp\alpha\sb{il} -
\delta\sb{il}f\sp\alpha\sb{kj})
\label{commute.gcalo}
\end{equation}
and all but the last of relations (\ref{commute.calo}). The dynamical
variables $(f\sb{ij})\sb{i,j=1,\ldots , N}$ are now
defined by  $f\sb{ij} = \sum\sb{\alpha =1}\sp M f\sb{ij}\sp\alpha$.
To see that the model (\ref{ham.calo})-(\ref{commute.gcalo}) is
integrable when $(f\sb{ii} = {\rm const})\sb{i=1,\ldots , N}$ we
construct the corresponding Lax matrix
\begin{equation}
L(\lambda) = \sum\sb{i=1}\sp N \left( p\sb i + \sum\sb{\alpha =1}\sp M
{f\sb{ii}\sp\alpha\over\lambda-\epsilon\sb\alpha}\right) e\sb{ii} +
\sum\sb{\stackrel{i,j =1}{i\neq j}}\sp N \left( {f\sb{ij}\over q\sb i
-q\sb j}+\sum\sb{\alpha =1}\sp M
{f\sb{ij}\sp\alpha\over\lambda-\epsilon\sb\alpha}\right) e\sb{ij},
\label{bigl}
\end{equation}
where $(e\sb{ij})\sb{kl} = \delta\sb{ik}\delta\sb{jl}$ and
$\epsilon\sb 1>\epsilon\sb 2>\cdots >\epsilon\sb M$ are
arbitrary parameters. This form of $L(\lambda)$ immediately reveals
that the generalisation of the Calogero-Moser model we discuss
here is achieved by  the coupling of the $M$-particle Gaudin model
\cite{gaudin1}  to the Calogero-Moser system.
Using equations (\ref{commute.calo}) and (\ref{commute.gcalo}) we can
write the Poisson brackets of the components of the Lax matrix
(\ref{bigl}) in the unified form
\cite{billey1}
\begin{equation}
\{L\sb 1(\lambda), L\sb 2(\mu)\} = [r\sb{12}(\lambda,\mu), L\sb
1(\lambda)+L\sb 2(\mu)] -  \sum\sb{\stackrel{i,j =1}{i\neq j}}\sp
N{f\sb{ii} - f\sb{jj}\over (q\sb i -q\sb j)\sp 2} e\sb{ij}\otimes
e\sb{ji}.
\label{lr}
\end{equation}
The r-matrix here is the dynamical one considered in \cite{billey1}
\begin{equation}
r\sb{12}(\lambda,\mu) = \sum\sb{\stackrel{i,j =1}{i\neq j}}\sp N \left(
{1\over\lambda -\mu} + {1\over q\sb i-q\sb j}\right) e\sb{ij}\otimes
e\sb{ji} +{1\over \lambda -\mu}\sum\sb i\sp N e\sb{ii}\otimes
e\sb{ii}.
\label{rmatrix}
\end{equation}
We note that  $r\sb{12}(\lambda ,\mu) =
-r\sb{21}(\mu ,\lambda)$.
{}From (\ref{lr}) we immediately learn that the model
(\ref{ham.calo}) is integrable provided $f\sb{ii} =f\sb{jj}$, $i,j =
1,\ldots ,N$, since in this case the Poisson bracket algebra
(\ref{lr}) takes the form (\ref{yang}). In particular, the model  is
integrable on the surfaces  $(f\sb{ii} =
{\rm const})\sb{i=1,\ldots , N}$. This final reduction is possible,
because each of $f\sb{ii}$ Poisson commutes with the Hamiltonian $H$. The
Hamiltonian $H$ is recovered as $H =\h \int\sb C {d\lambda\over 2\pi
i\lambda}\tr L\sp 2(\lambda)$, where $\lambda$ is considered  as a
complex variable and the contour $C$ encloses the origin.

\section{The N=2 Case}
Now  we focus on the $N=2$ case
for which we write a complete set of integrals of motion. In the
next section we will use the functional Bethe Ansatz procedure to separate
the variables in the Hamilton-Jacobi equations corresponding to these
integrals of motion.

In the centre of mass frame, the Lax matrix in this case has the form
(\ref{lax}), with
\begin{equation}
A(\lambda) =  p + \sum\sb{\alpha=1}\sp M {S\sb
3\sp\alpha\over\lambda-\epsilon\sb\alpha} ,\quad B(\lambda) = -{S\sb
-\over q} + \sum\sb{\alpha=1}\sp M {S\sb
-\sp\alpha\over\lambda-\epsilon\sb\alpha} , \quad C(\lambda) = {S\sb
+\over q} +\sum\sb{\alpha=1}\sp M {S\sb
+\sp\alpha\over\lambda-\epsilon\sb\alpha}
\label{2l}
\end{equation}
where $S\sb 3\sp\alpha$, $S\sb\pm\sp\alpha$ are generators of the
$so(2,1)$ Poisson algebra,
$$
\{S\sb 3\sp\alpha , S\sb\pm\sp\beta\} = \pm \delta\sb{\alpha
\beta}S\sb\pm\sp\beta ,\qquad \{S\sb +\sp\alpha , S\sb-\sp\beta\} =
2\delta\sb{\alpha \beta}S\sb 3\sp\beta
$$
and $(q,p)$ are the relative coordinates, $\{q,p\} =1 $, $q=q\sb 1
-q\sb 2$, $p=\h (p\sb 1-p\sb 2)$. The matrix (\ref{2l}) is obtained
from (\ref{bigl}) by subtracting the centre of mass motion. The
non-zero Poisson brackets of the
components of the Lax matrix (\ref{2l}) take the explicit form
\begin{eqnarray}
\{A(\lambda) , B(\mu)\} & = &{B(\lambda)-B(\mu)\over \lambda -\mu} +
{B(\lambda) \over q}, \nonumber \\
\{A(\lambda) , C(\mu)\} & = &-{C(\lambda)-C(\mu)\over \lambda -\mu} +
{C(\lambda) \over q}, \label{abc}\\
\{B(\lambda) , C(\mu)\} & = &2\left({A(\lambda)-A(\mu)\over \lambda
-\mu}  - {A(\lambda)-A(\mu)\over q}+{S\sb 3 \over q\sp 2} \right) .
\nonumber
\end{eqnarray}
The Poisson algebra (\ref{abc}) can be written in compact form
(\ref{yang}) with $r\sb{12}(\lambda,\mu)$ of the form (\ref{2r}), where
$$
a(\lambda,\mu) = {1\over{\lambda -\mu}}, \qquad b(\lambda,\mu) =
{1\over\lambda-\mu}+ {1\over q}, \qquad c(\lambda ,\mu)=
{1\over\lambda-\mu}- {1\over q},
$$
provided that $S\sb 3 =0$. This follows immediately from (\ref{lr}).
Writing explicitly the generating function (\ref{gt}),
$$
t(\lambda) =   p\sp 2 -  {S\sb- S\sb +\over q\sp 2} +  \sum\sb{\alpha
,\beta =1}\sp M {S\sb 3\sp\alpha S\sb 3\sp\beta + S\sb -\sp\alpha S\sb
+\sp\beta \over (\lambda-\epsilon\sb\alpha)(\lambda-\epsilon\sb\beta)}
+ \sum\sb{\alpha =1}\sp M {2pS\sb 3\sp\alpha +{1\over q}(S\sb
-\sp\alpha S\sb + - S\sb +\sp\alpha S\sb - )\over \lambda
-\epsilon\sb\alpha},
$$
we can compute that $\{t(\lambda), S\sb 3\} =0$, hence the reduction
$S\sb 3 =0$ can be performed. A complete set of integrals of motion
$H$, $G\sb\alpha$, $H\sb\alpha$, $\alpha =1,\ldots, M$ is found by
rewriting the generating function $t(\lambda)$ in the following form
\begin{equation}
t(\lambda) = H + \sum\sb{\alpha =1}\sp M {H\sb\alpha \over \lambda
-\epsilon\sb\alpha}  + \sum\sb{\alpha =1}\sp M {G\sb\alpha \over (\lambda
-\epsilon\sb\alpha)\sp 2},
\label{tcon}
\end{equation}
where
\begin{eqnarray}
H & = &p\sp 2 -
{S\sb- S\sb +\over q\sp 2}, \nonumber\\
G\sb\alpha& = & \left( S\sp\alpha\sb
3\right)\sp 2 + S\sb -\sp\alpha S\sb +\sp\alpha ,
\label{hamiltonians}\\
H\sb\alpha & = &\sum\sb{\stackrel{\beta =1}{\beta\neq\alpha}}\sp M
{2S\sb 3\sp\alpha
S\sb 3\sp\beta + (S\sb -\sp\alpha S\sb +\sp\beta + S\sb -\sp\beta
S\sb +\sp\alpha)\over \epsilon\sb\alpha -\epsilon\sb\beta}
+ 2pS\sb 3\sp\alpha +{1\over q}(S\sb -\sp\alpha S\sb + - S\sb
+\sp\alpha S\sb - ). \nonumber
\end{eqnarray}
Integrals of motion (\ref{hamiltonians}) are in involution provided that
$S\sb 3 = 0$. Notice that if this is the case, then  $\sum\sb{\alpha =1}\sp M
H\sb \alpha =0$. The first of  the integrals of
motion (\ref{hamiltonians}) is the
Hamiltonian of our system and for each $\alpha =1,\ldots ,M$,
$G\sb\alpha$  is a Casimir function hence it remains constant on each
symplectic leaf of the manifold on which the system is realised.
Therefore we have $M$ independent integrals of  motion.
We can represent the variables ${\bf S}\sp\alpha$ in
terms of the canonical coordinates and momenta $(x\sb\alpha ,
p\sb\alpha)$, $\{x\sb\alpha , p\sb\beta \} = \delta\sb{\alpha\beta}$,
$\alpha,\beta = 1,\ldots ,M$ as follows
\begin{equation}
S\sb 3\sp\alpha = \h x\sb\alpha p\sb\alpha , \quad S\sp\alpha\sb + =
\h p\sb\alpha\sp 2, \quad S\sb -\sp\alpha = -\h x\sb\alpha\sp 2.
\label{manifold}
\end{equation}
In the representation (\ref{manifold}) the  first integrals
(\ref{hamiltonians}) take the form
\begin{eqnarray}
H & = &  p\sp 2 +{R\sp 2\over 4q\sp 2}\sum\sb{\alpha =1}\sp M
p\sb\alpha\sp 2,\nonumber \\
G\sb\alpha  & = & 0 ,\qquad \alpha = 1,\ldots , M,   \label{hamiltonians1}\\
H\sb\alpha & = & -{1\over 4}\sum\sb{\stackrel{\beta
=1}{\beta\neq\alpha}}\sp M {M\sb{\alpha\beta}\sp 2\over
\epsilon\sb\alpha - \epsilon\sb\beta} + px\sb\alpha p\sb\alpha +
{1\over 4q}  \sum\sb{\stackrel{\beta
=1}{\beta\neq\alpha}}\sp M (p\sb\alpha\sp 2 x\sb\beta\sp 2
-x\sb\alpha\sp 2p\sb\beta\sp 2), \nonumber
\end{eqnarray}
where $R\sp 2 = \sum\sb{\alpha =1}\sp M x\sb\alpha\sp 2$ and
$M\sb{\alpha\beta} = p\sb\alpha x\sb\beta - x\sb\alpha p\sb\beta$. The
constraint $S\sb 3 = 0$ translates to
\begin{equation}
\sum\sb{\alpha =1}\sp M x\sb\alpha p\sb\alpha = 0.
\label{constraint}
\end{equation}
 If we choose $H$ as the Hamiltonian of the
system, as in fact  we are doing here,  then the constraint
(\ref{constraint}) implies that $R\sp 2={\rm  const}$ and we have
a two particle Calogero-Moser model coupled to the free motion on the
sphere $S\sp{M-1}$.

\section{Separation of Variables}
The dynamical system (\ref{hamiltonians1}) can be solved by
separation of variables in the Hamilton-Jacobi equations. The
separation is carried out in the framework of the functional Bethe
Ansatz \cite{sklyanin1} as described in the introduction. First we
consider the first of Eqs.(\ref{px}), which in our case reads
$$
{R\sp 2\over q} - \sum\sb{\alpha =1}\sp M {x\sb\alpha\sp 2\over
\lambda -\epsilon\sb\alpha} = 0.
$$
It is a polynomial equation of degree $M$ and it has $M$ different
solutions $X\sb i$, $i=1,\ldots ,M$. We use the Vieta theorem  to
derive the following expressions for $q$ and $x\sb\alpha\sp 2$,
$\alpha = 1,\ldots , M$
\begin{equation}
q= \sum\sb{i=1}\sp{M} X\sb i -  \sum\sb{\alpha =1}\sp M
\epsilon\sb\alpha ,\qquad
x\sb\alpha\sp 2 = {R\sp 2\over q}{\prod\sb{i=1}\sp{M} (X\sb i
-\epsilon\sb\alpha) \over \prod\sb{\stackrel{\beta
=1}{\beta\neq\alpha}}\sp M (\epsilon\sb\beta - \epsilon\sb\alpha)}.
\label{qx}
\end{equation}
We can choose the roots $X\sb i$, $i=1,\ldots , M$ in such a way
that $X\sb 1>\ldots >X\sb M$. {}From (\ref{qx}) we then learn that the
separation coordinates $X\sb i$, $i=1,\ldots , M$ satisfy the
inequalities
\begin{equation}
X\sb M<\epsilon\sb M< X\sb{M-1}<\ldots <X\sb 1<\epsilon\sb 1 \quad
{\rm or}\quad \epsilon\sb M<X\sb M<\ldots <\epsilon\sb 1<X\sb 1.
\label{range}
\end{equation}

Since $a(\lambda,\mu)$ obeys (\ref{condition}) it is clear that the
canonical momenta $P\sb i$, $i=1, \ldots ,M$ can be defined by the
second of Eqs.(\ref{px}). Therefore Eqs.(\ref{px}) give a full set of
canonical variables $(X\sb i , P\sb
i )\sb{i = 1,\ldots ,M}$, $\{X\sb i ,P\sb j
\} =\delta\sb{ij}$, and we can proceed to separation of variables.
We seek the common solution $S(X\sb 1,\ldots ,X\sb M)$ of the
Hamilton-Jacobi equations
$$
H\left({\partial S\over \partial X\sb 1},\ldots , {\partial S\over
\partial X\sb
M}, X\sb 1,\ldots ,X\sb M\right) = E, \quad H\sb\alpha\left({\partial
S\over \partial X\sb 1},\ldots , {\partial S\over \partial X\sb
M}, X\sb 1,\ldots ,X\sb M\right) = E\sb\alpha
$$
in the separated form $S(X\sb 1,\ldots ,X\sb M) = \sum\sb{i=1}\sp{M}
S\sb i (X\sb i )$. Here $E\sb\alpha$ are such that $\sum\sb{\alpha=1}\sp M
E\sb\alpha = 0$.  {}From (\ref{pt}) we find
the separated equations
\begin{equation}
 \left( {dS\sb i\over dX\sb i}\right)\sp 2 - E -\sum\sb{\alpha=1}\sp M
{E\sb\alpha\over X\sb i -\epsilon\sb\alpha} = 0, \qquad i=1,\ldots, M.
\label{sepa}
\end{equation}
If we consider only  a flow generated by the Hamiltonian $H$, then
$E\sb\alpha$, $\alpha =1,\ldots ,M$ have the meaning of arbitrary
separation constants.

The separated Hamilton-Jacobi equations (\ref{sepa}) can be also used
to express the Hamiltonian $H$ in terms of the separation coordinates
$(X\sb i , P\sb i )\sb{i = 1,\ldots ,M}$. Eliminating constants
$E\sb\alpha$, $\alpha =1,\ldots ,M$ from Eqs.(\ref{sepa}), we find that
$$
H = {\sum\sb{i=1}\sp M P\sb i\sp 2 \prod\sb{\alpha
=1}\sp M (X\sb i -\epsilon\sb\alpha) \prod\sb{\stackrel{j=1}{j\neq
i}}\sp M (X\sb i-X\sb j)\sp{-1}\over \sum\sb{i=1}\sp M \prod\sb{\alpha
=1}\sp M (X\sb i -\epsilon\sb\alpha) \prod\sb{\stackrel{j=1}{j\neq
i}}\sp M (X\sb i-X\sb j)\sp{-1}}.
$$

Finally we would like to stress that to derive separated
equations such as Eqs. (\ref{sepa}) we do not
have to specify to the representation (\ref{manifold}) of ${\bf
S}\sp\alpha$, $\alpha = 1,\ldots ,M$. In other words we can also
separate the Hamilton-Jacobi equations for integrals
(\ref{hamiltonians}). In this general case
the separated equations read
$$
 \left( {dS\sb i\over dX\sb i}\right)\sp 2 - E -\sum\sb{\alpha=1}\sp M
\left( {E\sb\alpha\over X\sb i -\epsilon\sb\alpha} + {G\sb\alpha\over
(X\sb i -\epsilon\sb\alpha)\sp 2}\right) = 0.
$$

\section{Quantisation}
The classical system (\ref{ham.calo})  can be quantised in the canonical
way by replacing the Poisson brackets
$\{ \;\; ,\;\;\}$ with the commutators $-i[\;\; ,\;\;]$. Also, the
functional Bethe Ansatz method can be quantised in this way
\cite{kuznetsov3}. We can
consider the Lax matrix (\ref{lax}) which generates the Gaudin algebra
(see e.g. \cite{brzezinski13})
\begin{equation}
[L\sb 1(\lambda), L\sb 2(\mu)] = i[r\sb{12}(\lambda,\mu), L\sb
1(\lambda)] -i[r\sb{21}(\mu ,\lambda),L\sb 2(\mu)].
\label{qyang}
\end{equation}
A generating function $t(\lambda)$ for commuting integrals of motion
is given by (\ref{gt}). We  consider r-matrices of the form
(\ref{2r}). Functions $a(\lambda ,\mu)$, $b(\lambda ,\mu)$, $c(\lambda
,\mu)$ may depend on dynamical variables, hence they are operator valued
functions in general. They satisfy similar conditions as before to
ensure that the commutator (\ref{qyang}) satisfies the Jacobi
identity.  In contrast to the classical case however, functions
$a(\lambda ,\mu)$, $b(\lambda ,\mu)$ and $c(\lambda
,\mu)$ must satisfy additional conditions to  make the
Gaudin algebra (\ref{qyang})
consistent. The sufficient ones are
\begin{eqnarray}
&& [a(\lambda ,\mu), A(\lambda)] = [a(\lambda ,\mu), B(\lambda)] =
[a(\lambda ,\mu), C(\lambda)] = 0 \nonumber ,\\
&& [b(\lambda ,\mu), B(\lambda)] =
[c(\lambda ,\mu), C(\lambda)] = 0. \label{add}
\end{eqnarray}
If the conditions (\ref{add})  are satisfied, then the algebra
(\ref{qyang}) takes the explicit form
\begin{eqnarray}
\left[ A(\lambda) , B(\mu)\right] & = &i\left( b(\lambda,\mu)B(\lambda)+
a(\mu,\lambda)B(\mu)\right) , \nonumber \\
\left[ A(\lambda) , C(\mu)\right] & = & -i\left( c(\lambda, \mu)C(\lambda)+
a(\mu,\lambda)C(\mu)\right) ,
\label{qgabc}\\
\left[ B(\lambda) , C(\mu)\right] & = &i\left(\left[ c(\lambda,\mu),
A(\lambda)\right]\sb+ +\left[ b(\mu,\lambda),A(\mu)\right]\sb+\right)  ,
\nonumber
\end{eqnarray}
where $[\;\; ,\;\;]\sb+$ denotes the anticommutator.

The separation of variables is then conducted in the way similar to
the classical case. We define  separation variables $(X\sb i, P\sb
i)$ by (\ref{px}). They are canonical if the condition
(\ref{condition}) is satisfied. In the quantum case however, Eqs.
(\ref{px}) are operator
equations, hence we have to specify the order of operators appearing
in $A(X\sb i)$ and $B(X\sb i)$. We always assume that the position
operators precede the momenta. We also assume that all substitutions
are done from the left. From this it follows that the new momenta
$P\sb i$ are not true observables since $P\sb i\neq P\sb i\sp\dagger$
in general. Now we show how the hermitian separation variables can be
defined and the separated Schr\"odinger equations obtained for the
Calogero-Moser-Gaudin model discussed in this paper.

We consider the $N=2$ case and we take the L-operator (\ref{2l}) as a
Lax matrix. We observe that the Gaudin algebra  for this
model is obtained directly from (\ref{abc}) by replacing the Poisson
brackets $\{ \;\; ,\;\;\}$ with  the commutators $-i[\;\; ,\;\;]$. From
definition (\ref{gt}) we find a quantum
generating function for  commuting integrals of motion
$$
t(\lambda)=   p\sp 2 \!-\! {[S\sb-, S\sb+]\sb+ \over 2q\sp 2} +
\sum\sb{\alpha
,\beta =1}\sp M {S\sb 3\sp\alpha S\sb 3\sp\beta + \h [S\sb -\sp\alpha ,S\sb
+\sp\beta]\sb + \over (\lambda-\epsilon\sb\alpha)(\lambda-\epsilon\sb\beta)}
+ \sum\sb{\alpha =1}\sp M {2pS\sb 3\sp\alpha +{1\over 2q}\left( [S\sb
-\sp\alpha ,S\sb +]\sb+ - [S\sb +\sp\alpha S\sb - ]\sb+\right)\over \lambda
-\epsilon\sb\alpha}.
$$
The full set of  conserved quantities  reads
\begin{eqnarray}
H & = &p\sp 2 -
{[S\sb-, S\sb +]\sb+\over 2q\sp 2}, \nonumber\\
G\sb\alpha& = &\left( S\sp\alpha\sb
3\right)\sp 2 + \h[S\sb -\sp\alpha ,S\sb +\sp\alpha]\sb+,
\label{qhamiltonians}\\
H\sb\alpha & = &\sum\sb{\stackrel{\beta =1}{\beta\neq\alpha}}\sp M
{2S\sb 3\sp\alpha
S\sb 3\sp\beta + (S\sb -\sp\alpha S\sb +\sp\beta  + S\sb -\sp\beta
S\sb +\sp\alpha)\over \epsilon\sb\alpha -\epsilon\sb\beta}
+ 2pS\sb 3\sp\alpha +{1\over 2q}\left( [S\sb -\sp\alpha ,S\sb +]\sb+ - [S\sb
+\sp\alpha, S\sb -]\sb+ \right) , \nonumber
\end{eqnarray}
and $t(\lambda)$ is given by (\ref{tcon}). Similarly to the classical
case, the first of the integrals of motion
(\ref{qhamiltonians}) plays the role of the Hamiltonian of the quantum
system and $\sum\sb{\alpha =1}\sp MH\sb\alpha =0$. For each $\alpha
=1,\ldots ,M$, the integral $G\sb\alpha$ is a quadratic Casimir
operator of the algebra $so\sb\alpha(2,1)$ generated by ${\bf
S}\sp\alpha$. Hence each of $G\sb\alpha$ is a number in any
irreducible representation of the system. The particular
representation is obtained by realising   spin variables ${\bf
S}\sp\alpha$, $\alpha =1,\ldots ,M$ in
terms of  the canonical variables $(x\sb\alpha ,
p\sb\alpha)$, $[x\sb\alpha , p\sb\beta ] =
i\delta\sb{\alpha\beta}$,
$\alpha,\beta = 1,\ldots ,M$ as follows:
\begin{equation}
S\sb 3\sp\alpha = \q (x\sb\alpha p\sb\alpha + p\sb\alpha x\sb\alpha) ,
\quad S\sp\alpha\sb + =
\h p\sb\alpha\sp 2, \quad S\sb -\sp\alpha = -\h x\sb\alpha\sp 2.
\label{qmanifold}
\end{equation}
In the representation (\ref{qmanifold}) the first integrals
(\ref{qhamiltonians}) take the form (compare
Eqs.(\ref{hamiltonians1}))
\begin{eqnarray}
H & = &  p\sp 2 +{R\sp 2\over 4q\sp 2}\sum\sb{\alpha =1}\sp M
p\sb\alpha\sp 2,\nonumber \\
G\sb\alpha & = & {3\over 16} , \qquad \alpha
= 1,\ldots ,M, \nonumber \\
H\sb\alpha  &=& -{1\over 4}\sum\sb{\stackrel{\beta
=1}{\beta\neq\alpha}}\sp M {M\sb{\alpha\beta}\sp 2+1/2\over
\epsilon\sb\alpha - \epsilon\sb\beta} + \h p[x\sb\alpha,
p\sb\alpha]\sb + +
{1\over 4q}  \sum\sb{\stackrel{\beta
=1}{\beta\neq\alpha}}\sp M (p\sb\alpha\sp 2 x\sb\beta\sp 2
-x\sb\alpha\sp 2p\sb\beta\sp 2), \nonumber
\end{eqnarray}
Notice that $R$ is an operator now. We can proceed to separation of
variables in the Schr\"odinger equations, defining new canonical
coordinates by (\ref{px}). As we observed, the momenta $P\sb i$,
$i=1,\ldots ,M$ are
not hermitian, therefore true separation momenta are still to be
defined. We use Eqs.(\ref{qx}) to find that
$$
B(\lambda) = {R\sp 2\over 2q} {\prod\sb{i=1}\sp M (\lambda -X\sb
i)\over \prod\sb{\alpha =1}\sp M(\lambda-\epsilon\sb\alpha)}, \qquad
A(\lambda) = {2q\over R\sp 2}B(\lambda)\left( p + q\sum\sb{i =1}\sp
M{1\over\lambda -X\sb i}D\sb i P\sb i\right),
$$
where
$$
D\sb i =  {\prod\sb{\alpha =1}\sp M (X\sb i -\epsilon\sb\alpha)
\over q\prod\sb{\stackrel{j=1}{j\neq i}}\sp M(X\sb i -X\sb j)},
\qquad i=1,\ldots ,M.
$$
The expression for $A(\lambda)$ is derived by the analysis of the
behaviour of $A(\lambda)$ at $\lambda = X\sb 1, \ldots ,X\sb M$ and
$\lambda = \infty$. {}From (\ref{range}) we learn that each $D\sb i$
is positive. Using the fact that $[X\sb i, P\sb j] =
i\delta\sb{ij}$ and also that
$$
\sum\sb{i=1}\sp M {1\over\lambda -X\sb i}{\prod\sb{\alpha =1}\sp M
(X\sb i -\epsilon\sb\alpha) \over \prod\sb{\stackrel{j=1}{j\neq i}}\sp
M(X\sb i -X\sb j)} = {\prod\sb{\alpha =1}\sp M
(\lambda -\epsilon\sb\alpha) \over \prod\sb{i=1}\sp
M(\lambda  -X\sb i)} -1,
$$
we find that
$$
A\sp\dagger(\lambda) = {2q\over R\sp 2}B(\lambda)\left( p + q\sum\sb{i =1}\sp
M{1\over\lambda -X\sb i}P\sp\dagger\sb i D\sb i\right).
$$
Thus
\begin{equation}
D\sb i P\sb i = P\sb i\sp\dagger D\sb i,
\end{equation}
because $A\sp\dagger(\lambda) = A(\lambda)$.
Therefore we can define hermitian operators
\begin{equation}
\Pi\sb i \equiv \sqrt{D\sb i} P\sb i {1\over \sqrt{D\sb i}}.
\label{pi}
\end{equation}
The operators (\ref{pi}) are canonically conjugate to $X\sb i$ and
play the role of true separation momenta. Directly from the definition of
the $P\sb i$ we can derive  equations of motion (\ref{pt}), which
in terms of the $\Pi\sb i$ read
\begin{equation}
{1\over \sqrt{D\sb i}}\Pi\sb i\sp 2\sqrt{D\sb i} \Psi\left( X\sb
1,\ldots , X\sb M\right) \!- \!\left( E\!
+\!\sum\sb{\alpha=1}\sp M \left(
{E\sb\alpha\over X\sb i\! -\! \epsilon\sb\alpha}\! +\! {3/16\over (X\sb
i\!- \! \epsilon\sb\alpha)\sp 2}\right)\right)\Psi\left( X\sb
1,\ldots , X\sb M\right) = 0, \
\label{schr}
\end{equation}
$i=1,\ldots, M$. Here $E,E\sb\alpha$ are  eigenvalues of the operators
$H,H\sb\alpha$, $\alpha =1,\ldots ,M$. To see that Eqs.(\ref{schr})
are really separation equations for the model we set
\begin{equation}
\Psi(X\sb 1, \ldots ,X\sb M) = \sqrt{|qV|}\prod\sb{i=1}\sp M \Psi\sb
i(X\sb i),
\label{psi}
\end{equation}
where $V$ denotes the Vandermonde determinant $V = \prod\sb{i<j}\sp M
(X\sb i -X\sb j)$.
Inserting the wave function (\ref{psi}) into equations (\ref{schr})
and representing $\Pi\sb i$ by $-i{d\over dX\sb i}$, $i =1,\ldots ,M$,
we obtain
\begin{equation}
{1\over \sqrt{C\sb i}}{d\sp 2\over dX\sb i\sp 2}\left(\sqrt{C\sb i}
\Psi\sb i\right) + E\Psi\sb i
+\sum\sb{\alpha=1}\sp M \left(
{E\sb\alpha\over X\sb i -\epsilon\sb\alpha} + {3/16\over (X\sb
i-\epsilon\sb\alpha)\sp 2}\right)\Psi\sb i = 0, \quad i =1,\ldots, M,
\label{schr.sep}
\end{equation}
where
$$
C\sb i = \left| \prod\sb{\alpha =1}\sp M (X\sb i -\epsilon\sb
\alpha)\right|.
$$
Equations (\ref{schr.sep}) are then the separated Schr\"odinger
equations for the generalised Calogero-Moser model. Finally we notice
that the separated equations in the
general representations of the algebras $so\sb\alpha(2,1)$ can be
obtained from (\ref{schr.sep}) by replacing $3/16$ with the
eigenvalues $g\sb\alpha$ of $G\sb\alpha$ which characterise these
representations, i.e.
$$
{1\over \sqrt{C\sb i}}{d\sp 2\over dX\sb i\sp 2}\left(\sqrt{C\sb i}
\Psi\sb i\right) + E\Psi\sb i
+\sum\sb{\alpha=1}\sp M \left(
{E\sb\alpha\over X\sb i -\epsilon\sb\alpha} + {g\sb\alpha\over (X\sb
i-\epsilon\sb\alpha)\sp 2}\right)\Psi\sb i = 0, \quad i =1,\ldots, M.
$$

\section{Conclusion}
In this paper we have described an integrable
generalisation of the Calogero-Moser model, which is
achieved by coupling the  $M$-particle Gaudin system to the
Calogero-Moser model. The
integrability of the model has been shown
by  using the recently introduced dynamical r-matrix
\cite{billey1}. We have also shown that  the functional Bethe-Ansatz
of \cite{sklyanin1} can be  employed to separate the variables in
 models governed by this r-matrix despite its dependence on
dynamical variables. The situation is similar to the one discussed
in \cite{eilbeck1}. We have used this
separation of variables procedure in the  case in which there are
two Calogero-Moser particles. We have also used the quantum
counterpart of this procedure \cite{kuznetsov3} to separate
variables in the Schr\"odinger equations.

The model discussed in this paper is simply an example of a system
which is governed by a dynamical r-matrix and can be solved by
separation of variables in the framework of the functional Bethe
Ansatz.  By an analogy to the non-dynamical r-matrix
case, one can expect that there are several other models which are
governed by dynamical r-matrices
of the type discussed here, and hence can be solved in a described way.
This observation opens up new possibilities for constructing and
solving integrable models as well as for getting a deeper insight into
the nature of the models governed by dynamical r-matrices.

\newpage

\noindent {\bf ACKNOWLEDGEMENT}

\noindent I would like to thank Alan Macfarlane for encouragement and
discussions.

\end{document}